\documentclass[fleqn,10pt]{wlscirep}
\usepackage{setspace}
\usepackage{textcomp}
\title{Phonon-assisted two-photon interference from remote quantum emitters}

\author[1,*]{Marcus Reindl}
\author[2,*]{Klaus D. Jöns}
\author[1]{Daniel Huber}
\author[1]{Christian Schimpf}
\author[3,1,4]{Yongheng Huo}
\author[2]{Val Zwiller}
\author[1]{Armando Rastelli}
\author[1,*]{Rinaldo Trotta}

\affil[1]{Institute of Semiconductor and Solid State Physics, Johannes Kepler University Linz, 4040, Austria}
\affil[2]{Department of Applied Physics, Royal Institute of Technology Stockholm, 106 91, Sweden}
\affil[3]{Institute for Integrative Nanosciences, IFW Dresden, 01069, Germany}
\affil[4]{Hefei National Laboratory for Physical Sciences at Microscale, University of Science and Technology Shanghai, 201315, China}

\keywords{Quantum dots, Quantum optics, Two-photon interference, Entanglement, Resonant two-photon excitation}

\begin{abstract}
Photonic quantum technologies~\cite{OBrien2010} are on the verge of finding applications in everyday life with quantum cryptography~\cite{Gisin.Ribordy.ea:2002} and the quantum internet~\cite{Kimble:2008} on the horizon. Extensive research has been carried out to determine suitable quantum emitters~\cite{Aharonovich.Englund.ea:2016} and single epitaxial quantum dots are emerging as near-optimal sources~\cite{Somaschi.Giesz.ea:2016,Ding.He.ea:2016} of bright~\cite{Claudon2010}, on-demand, highly indistinguishable~\cite{He2013} single photons and entangled photon pairs\cite{Muller.Bounouar.ea:2014,Trotta2014}. In order to build up quantum networks, it is now essential to interface remote quantum emitters. However, this is still an outstanding challenge, as the quantum states of dissimilar ”artificial atoms” have to be prepared on-demand with high fidelity, and the generated photons have to be made indistinguishable in all possible degrees of freedom~\cite{Flagg.Muller.ea:2010,Patel.Bennett.ea:2010,Gold.Thoma.ea:2014,Giesz.Portalupi.ea:2015}. Here, we overcome this major obstacle and show an unprecedented two-photon interference (visibility of 51$\pm$5\,\%) from remote strain-tunable GaAs quantum dots ~\cite{Kumar2011,Huo2013} emitting on-demand photon-pairs. We achieve this result by exploiting for the first time the full potential of the novel phonon-assisted two-photon excitation scheme~\cite{Glassl2013}, which allows for the generation of highly indistinguishable (visibility of 71$\pm$9\,\%) entangled photon-pairs (fidelity of 90$\pm$2\,\%), it enables push-to button biexciton state preparation (fidelity of 80$\pm$2\,\%) and it outperforms conventional resonant two-photon excitation schemes in terms of robustness against environmental decoherence. Our results mark an important milestone for the practical realization of quantum repeaters and complex multi-photon entanglement experiments involving dissimilar artificial atoms.
\end{abstract}
\begin{document}

\flushbottom
\maketitle

\thispagestyle{empty}

\section*{Introduction}

One of the very first requirements to observe ideal on-demand single photon emission is the population inversion of the quantum emitter's excited state. Such a preparation of the quantum state is usually achieved via coherent excitation using resonant laser pulses, leading to an inverted two-level system performing Rabi oscillations~\cite{Stievater2001}. While single pulse resonant excitation of a quantum dot~(QD) has been used to achieve high state preparation fidelities and remarkable single photon properties, this scheme cannot be used to efficiently prepare the biexciton state, the key step to achieve polarization-entangled photon-pairs generation with QDs. This task can be instead accomplished using two-photon excitation (TPE)~\cite{Brunner.Abstreiter.ea:1994,Stufler.Machnikowski.ea:2006,Jayakumar.Predojevic.ea:2013} techniques which have recently led to the generation of on-demand entangled photon pairs~\cite{Muller.Bounouar.ea:2014,Huber2016}. This coherent excitation scheme, however, has one important drawback. Small fluctuations in the laser pulse area or energy as well as fluctuations in the QD environment result in a strong variation of the excited state population probability that, in turn, affects the efficiency of photon generation. Envisioned quantum communication applications demand instead for more robust excitation schemes, being immune against these sources of "environmental decoherence" and ensuring on-demand generation of single and entangled photon-pairs.
In principle, it is possible to overcome these problems by taking advantage of the solid state nature of QDs, and in particular of their coupling to acoustic phonons. Despite the phonon-assisted excitation scheme is inherently incoherent, it has been proposed~\cite{Glassl2013} and recently demonstrated \cite{Ardelt.Hanschke.ea:2014, Quilter.Brash.ea:2015,Bounouar.Muller.ea:2015} that population inversion of X and XX states coupled to a quasicontinuum of vibrational modes is indeed possible. 
Nonetheless, the capability of this technique to generate highly indistinguishable single and entangled photons has not been explored so far. In this letter, we show for the first time that phonon-assisted two-photon excitation of QDs allows for the generation of highly indistinguishable entangled photon-pairs. In comparison with standard excitation schemes, we demonstrate that this method is more resilient against environmental decoherence limiting the XX or X preparation fidelity in conventional TPE schemes. Most importantly, we exploit its addressability with a wide-range of laser detunings to prepare on demand two remote and dissimilar QDs and to let the generated photons interfere at a beam splitter, a key experiment for the realization of an all-optical quantum repeater~\cite{Duan.Lukin.ea:2001,Sangouard.Simon.ea:2011}.

\section*{Results and Discussion}
\begin{figure}[ht]
\centering
\includegraphics{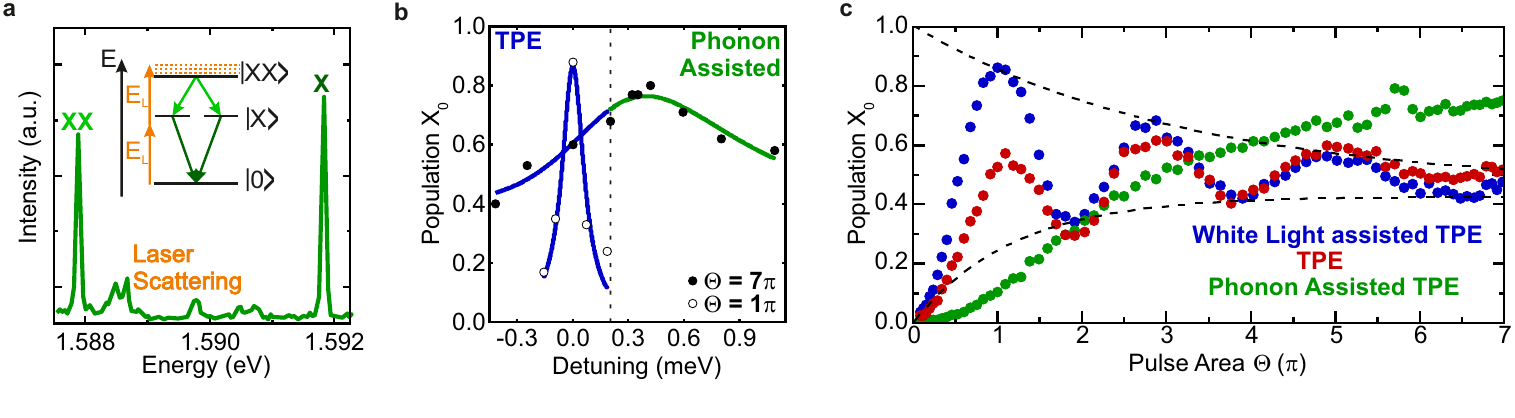}
\caption{\textbf{Spectrum and power dependent studies.} (a) Spectrum of the QD under phonon-assisted TPE for optimized detuning of the laser energy and pulse length ($\mathrm{E_L=1.5901}$\,eV, $\mathrm{\tau_p=10}$\,ps). X and XX are clearly visible and the residual lines are attributed to laser scattering  as well as weakly excited charged states (Inset: Phonon-assisted TPE scheme where the laser pulses are tuned to half the energy of the vibrational quasicontinuum (dashed-orange) coupled to the biexcition state). (b) Population of the X state as a function of the laser detuning for varying excitation power. While the traditional TPE (blue) suffers from a steep drop in inversion efficiency, a stable plateau can be observed exploiting the QDs phonon sideband (green). The measured data for high excitation powers (7$\pi$) are interpolated with an asymmetric-modulated Gaussian function while the low power data ($\pi$-pulse) is fitted as a Lorentzian function.  (c) Power dependent studies of the resonant TPE with (blue) and without (red) white light. The envelope of the Rabi oscillation is modeled with a single exponential damping. The results of the phonon-assisted excitation scheme are shown as green circles.}
\label{fig:figure1}
\end{figure}

\begin{figure}[ht]
\centering
\includegraphics{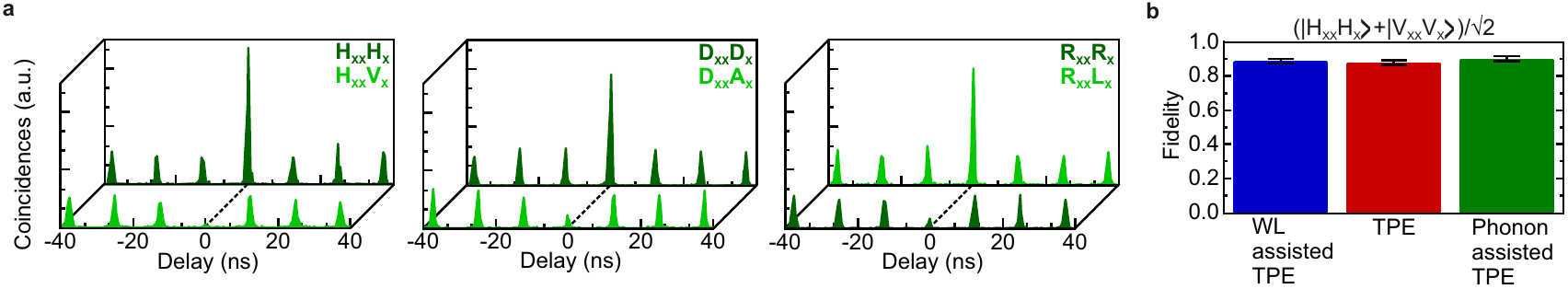}
\caption{\textbf{Comparison of entanglement.} (a) XX-X cross-correlation measurements under phonon-assisted TPE for different polarization detection bases: rectilinear (H,V), diagonal (D,A) and circular basis (R,L). (b) Fidelity to the expected Bell state for all the different excitations methods.}
\label{fig:figure2}
\end{figure}

\begin{figure}[h]
\centering
\includegraphics{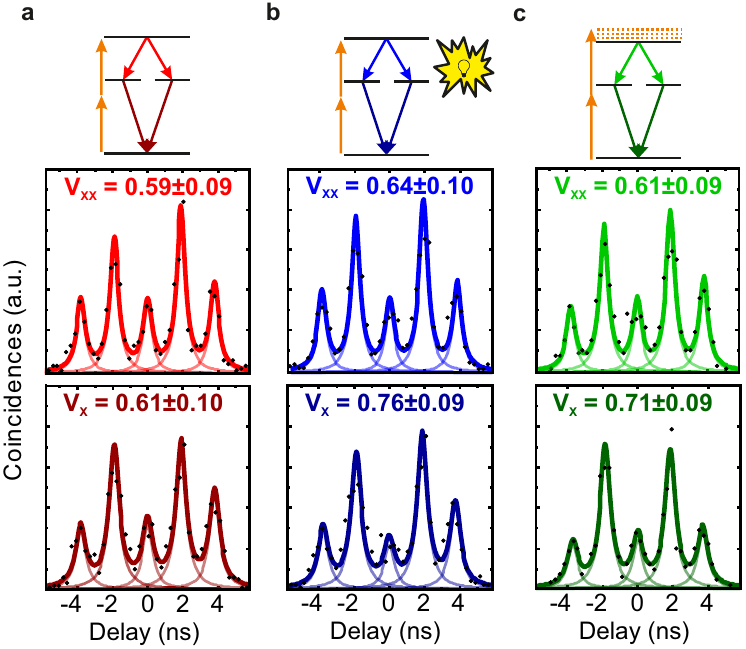}
\caption{\textbf{Single QD two-photon interference.} The two-photon interference is performed on the same QD for (a) the standard TPE (b) the white light assisted TPE and (c) the phonon-assisted TPE, as schematically illustrated on top of each panel. The histograms envelope function (bold) is the sum of 5 Lorentzian peaks fitted to the Hong-Ou-Mandel quintuplet. The resulting interference visibilities are reported in each panel.}
\label{fig:figure3}
\end{figure}

\begin{figure}[ht]
\centering
\includegraphics{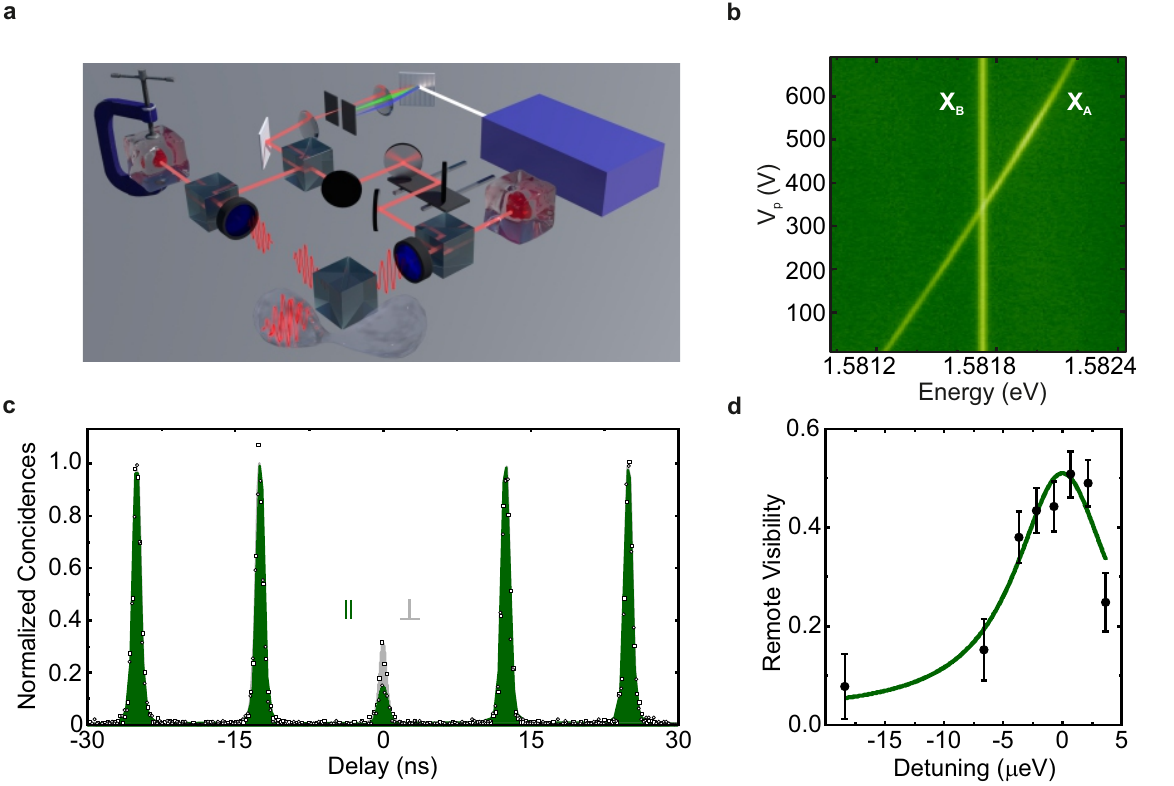}
\caption{\textbf{Two-photon interference from remote QDs.} (a) Illustration of the interference of remote GaAs QD single photon sources (ice cubes). A pulse-shaped laser is split and excites both QDs via the phonon-assisted TPE. Strain tuning (c-clamp) of one QD allows for the precise frequency matching of the emitted photons. (b) Sweep of the piezo voltage ($\mathrm{V_p}$) on QD A to achieve color coincidence with the X from QD B. (c) Second-order correlation for remote X photons excited in the phonon-assisted TPE scheme in co- and cross- configuration. (d) The visibility of the remote two-photon interference as a function of detuning between the two X photons.}
\label{fig:figure4}
\end{figure}

We focus our study on highly symmetric GaAs/AlGaAs QDs obtained via the droplet-etching method \cite{Huo2013} (see supporting note 1). The photon pairs emitted from this specific type of QDs have recently shown unprecedented high degree of entanglement as well as indistinguishability~\cite{Huber2016}.
A typical spectrum of our highly symmetric GaAs/AlGaAs QDs under phonon-assisted excitation is shown in Fig.\ref{fig:figure1}(a). To address the vibrational modes coupled to the XX state, the excitation laser is blue detuned by $\mathrm{\Delta}$ from the two-photon resonant case ($\mathrm{\Delta=0}$\,meV) towards the X transition  (Fig.\ref{fig:figure1}(a)). The best excitation parameters for optimum state preparation are inherently locked to the materials deformation potential and QD structural details, which determines the coupling of the exciton complexes to the acoustic phonons of the solid state environment \cite{Krummheuer2002}, thus leading to the excitonic phonon sidebands~\cite{Besombes2001}. The optimal detuning energy for the investigated type of QD system is around $\mathrm{\Delta=0.4}$\,meV for a pulse length of $\mathrm{\tau_p=10}$\,ps (see supporting note 2). It is important to point out that the laser energy can be swept across a range of 0.2\,meV without perturbing the state preparation fidelity (5\% population change, see supporting note 2), while in the conventional TPE the population varies for more than 80\% on the same energetic range taking into account $\pi$-pulse excitation (Fig.\ref{fig:figure1}(b)). The stability in the preparation fidelity offered by the phonon-assisted scheme is particularly relevant for this work and will be later used to address remote QDs with the same excitation laser. Before examining this point in more detail, we first discuss the robust nature of the phonon-assisted scheme in comparison with the standard TPE.  
\\
We first study the power dependence of the standard resonant TPE in the same excitation conditions and on the same QD (Fig.\ref{fig:figure1}(c) red curve). Interestingly, the TPE manifests itself as oscillations of the state occupation probability locked to $\frac{1}{2}$. While a possible explanation of this effect is the presence of a chirped laser pulse in conjunction with phonon-induced damping \cite{Reiter2012,Glassl2013}, we show that it is instead connected to the details of the QD environment: The power dependence changes considerably as we additionally illuminate the QD with a weak white-light source (Fig.\ref{fig:figure1}(c) blue curve) revealing traditional phonon-damped Rabi oscillations \cite{Forstner2003,Ramsay2010} with state population as high as 88$\pm$2\,\%. We attribute these modifications (which are particularly pronounced at the $\pi$-pulse) to saturation of crystal defects located in the vicinity of the QD. In the absence of the white light, these defects release/trap charge carriers, thus giving rise to a fluctuating electric field \cite{Kamada2008}. We hypothesize that the white light not only stabilizes the electric field experienced by the QD~\cite{Gazzano.MichaelisdeVasconcellos.ea:2013} (see below) but also suppresses/saturates recombination channels (probably charged XX states) which hamper the radiative recombination of the XX into the X state. Obviously, for high values of the pulse area, the effect becomes negligible as the carrier-phonon interaction dominates.
While the effect of the white light on the Rabi oscillations has never been reported so far, the addition of off-resonant lasers is common practice in experiments performed with QDs driven resonantly~\cite{Jahn.Munsch.ea:2015,Bennett.Lee.ea:2016}, even if the exact origin of its effect is not yet fully understood. We also point out that the effect of the white light on the QD driven with a $\pi$-pulse differs from QD to QD, with an increase in the population probability which ranges from 10\% to 50\% (see Fig.\ref{fig:figure1}(c)). Moreover, the intensity of the white light needed to achieve the optimal conditions is also QD-specific. Thus, the TPE schemes are not ideal for applications in complex networks and experiments with multiple sources. In stark contrast, no remarkable effect of the white light can be observed under phonon-assisted resonant excitation (\textless 5\% change in preparation fidelity), probably due to the large laser power needed, which also stabilizes the environment. Another important advantage of the phonon-assisted scheme is that it is inherently immune to fluctuations of the laser pulse area (see Fig.\ref{fig:figure1}(c)) due to its incoherent nature. More specifically, when the state preparation fidelity is maximum, a 10\% fluctuation of the pulse area leads to a negligible (\textless 1\%) change in the state population. In the standard TPE, the same fluctuation of pulse area gives rise to at least 7\% variation in the state fidelity. Finally, we emphasize that the phonon-assisted scheme allows preparing the excited state with very high fidelity, which is as high as 80$\pm$2\,\% for the highest laser pulse area available.
\\
After demonstration of the robust nature of the phonon assisted excitation scheme, we now investigate the quality of the generated photons in terms of entanglement fidelity and photon indistinguishability. We start out measuring the fidelity to the maximally entangled Bell state (see supporting note 1 and 4) using a QD with small FSS (1.3$\pm$0.5$\,\mu$eV). The polarization-resolved XX-X cross-correlation measurements used to estimate the fidelity are shown in Fig.\ref{fig:figure2}(a) under phonon-assisted excitation. These data yields a fidelity of f = 90$\pm$2\,\% which is identical (within the experimental error) to the values obtained when the QD is driven under strict TPE (with and without white light, Fig.\ref{fig:figure2}(b)). Therefore, the three different excitation schemes give rise to the same level of entanglement of the emitted photons. This is an expected result, as the fidelity is determined by three main contributions: (i) the relative value of the FSS with respect to the natural linewidth \cite{Stevenson2008}, (ii) recapture processes \cite{Dousse2010,2015arXiv151003897J} increasing the multiphoton emission and (iii) the hyperfine interaction \cite{Deng2006,Chekhovich2013}.
Since the lifetime (i) as well as the single photon purity (ii) and the hyperfine interaction (iii) are not affected by the excitation scheme (see supporting note 3), the fidelity to the Bell state is predicted to remain constant, as indeed measured experimentally (Fig.\ref{fig:figure2}(b)).
\\
The different excitation methods are instead expected to have a pronounced role in the indistinguishability of consecutive photons emitted by the same QD, as measured in an Hong-Ou-Mandel type experiment~\cite{Hong.Ou.ea:1987} on the XX and X photons. The time delay between consecutive photons was set to 2\,ns via a Mach-Zehnder interferometer in the excitation path. The observation of the typical two-photon interference quintuplet is presented in Fig.\ref{fig:figure3}, together with the visibility values $\mathrm{V_{X}}$ and $\mathrm{V_{XX}}$, which are calculated taking into account the imperfections of the interference beam splitter (see supporting note 1 and 5).
If we first take a look at the standard TPE (Fig.\ref{fig:figure3}(a)), we observe the usual tendency of XX and X photons with visibilities around 60\%. The stabilization of the QD environment throughout illumination with the white light source (Fig.\ref{fig:figure3}(b)), however, leads to an evident (slight) increase of the X (XX) visibility. This is reasonable as the X is more sensitive to spectral diffusion mediated by temporally charged defects \cite{Kamada2008} than the screened potential of the fully occupied XX state. The weaker visibility of the XX transition, on the other hand, can be related to the XX probing an extraordinary noise environment \cite{Kuhlmann2013} and/or suffers from an initially higher pure dephasing rate\cite{Vagov2003}. Most importantly, under phonon-assisted excitation of the two-level system (Fig.\ref{fig:figure3}(c)) a remarkably high level of indistinguishability (comparable to the TPE under white light illumination) can be observed. This demonstrates that the time jitter introduced by phonon relaxation is negligible in our measured values of photon indistinguishabilities. To summarize, the phonon-assisted two-photon excitation scheme not only leads to the generation of highly indistinguishable entangled photon-pairs but it is more robust than the standard two-photon excitation schemes. Yet, this scheme has an additional elegant advantage: it allows performing two-photon interference between remote QDs driven by the same pulsed laser of locked frequency. This is a direct consequence of the wide range of laser detunings that allows to achieve the maximum population inversion and it is in contrast to the traditional two-photon excitation schemes, which instead require a precise control of the laser energy for each individual QD featuring dissimilar XX binding energies. The phonon-assisted two-photon excitation is instead a universal clocked excitation for arbitrary large numbers of QDs, a scalable approach for quantum optics.
\\
The excitation of the remote emitters is timed so that the individual single photons from the two QDs, depicted as ice cubes in Fig.\ref{fig:figure4}(a), overlap on a beam splitter performing two-photon interference. For this experiment, we fabricate a second sample prepared on top of a piezoelectric actuator to provide tunability of the QD emission lines \cite{Trotta2012} (see methods) and to ensure frequency matching of the photons impinging at the beam splitter. Two X transitions with almost identical lifetimes and high single photon purity from two remote QDs (see supporting note 6) are tuned to the exact same frequency by applying a voltage across the piezoelectric actuator mounted below QD\,A (Fig.\ref{fig:figure4}(b)). In this condition, the difference in the biexciton binding energy between the two QDs is around 0.1\,meV. It implies that under standard TPE, the optimal state inversion for the two QDs cannot be achieved with the same laser (see the discussion above). In contrast, under phonon-assisted TPE we can prepare the two QDs with the highest probability by simply finding the optimum laser detuning for both, which in this special case turns out to be $\mathrm{\Delta=0.36}$\,meV (Fig.\ref{fig:figure1}(b)). The resulting correlation measurements are depicted in Fig.\ref{fig:figure4}(c) for the cross- and co-polarized configurations.
We would like to point out that in contrast to the two-photon interference of consecutive photons from a single source - where the cross-polarized configuration always yields a value of $\mathrm{g_{\perp}^{(2)}(0)=0.5}$ - this condition is not necessarily realized when combining remote single photon sources. In particular, long time-scale blinking reduces the value of $\mathrm{g_{\perp}^{(2)}(0)}$ even when the average intensities of the two emitters are kept the same~\cite{Jons2017}. Thus, it is crucial to first determine the cross-polarized correlation and to evaluate the real remote two-photon interference visibility $\mathrm{V_{remote}}$ as follows:
\begin{equation}
\mathrm{V_{remote}=\frac{   g_{\perp}^{(2)}(0) - g_{\parallel}^{(2)}(0)   }{  g_{\perp}^{(2)}(0)} }
\end{equation}
The optimized value for the overlap of the individual photon energies is then found by sweeping the X transition of QD\,A in steps of voltages that modify the energy on the order of a fraction of the linewidth, as demonstrated in Fig.\ref{fig:figure4}(d). By doing so we report on a remote interference visibility as high as $\mathrm{V_{remote}=51}$$\pm$5\,\%, one of the highest value ever observed for QDs without the need of any temporal/spectral selection. So far only coherently scattered~\cite{Gao.Fallahi.ea:2013} or Raman photons~\cite{He.He.ea:2013} from remote QDs achieved higher two-photon interference visibilities. However, these excitation schemes cannot be used to generate pairs of photons and does not allow for the on-demand state preparation, both important prerequisites for quantum relays based on entangled photons from QDs ~\cite{Trotta20152}. The visibility is in good agreement with the theoretical limits obtained from Michelson interferometry (see supporting note 6). Currently we are only limited by the non Fourier-limited photon emission of our QDs. A possible way to overcome this problem is to use devices that enable the application of electric fields ~\cite{Kuhlmann2015} and/or photonic cavities to shorten the lifetime of the transitions via the Purcell effect.
\\
In conclusion, we performed two-photon interference between photons emitted by two remote QDs with a visibility of 51$\pm$5\,\%, by using the full power of the novel phonon-assisted two-photon excitation scheme. In a comprehensive study, we compare different resonant excitation schemes and show that the phonon-assisted state preparation is a robust method to generate on-demand single pairs of highly entangled and indistinguishable photons from semiconductor quantum dots. Our results paves the way towards entanglement swapping experiments between independent QDs and other complex multi-source experiments.

\section*{Associated Content}
\subsection*{Supporting Information}

Methods, Detuning parameters of the phonon-assisted TPE; lifetime and auto-corrleation; fidelity evaluation; two-photon interference using the same QD; two-photon interference from remote QDs.

\section*{Author Information}
\subsection*{Corresponding Authors}
*(M.R.) E-mail: marcus.reindl@jku.at \\
*(K.D.J.) E-mail: klausj@kth.se \\
*(R.T.) E-mail: rinaldo.trotta@jku.at

\subsection*{Author contributions}
M.R.,K.D.J. and D.H performed the measurements with the help from C.S. and R.T.. M.R. made the data analysis with the help from K.D.J.,D.H. and R.T.. Y.H. grew the sample with support of A.R.. C.S. processed the sample. M.R., K.D.J. and R.T. wrote the manuscript with the help from all the authors. R.T. conceived the experiment and supervised the project.

\subsection*{Notes}
The authors declare no competing financial interests.

\section*{Acknowledgements}
This work was financially supported by the ERC Starting Grant No. 679183 (SPQRel) and European Union Seventh Framework Programme (FP7/2007-2013) under grant agreement no. 601126 (HANAS). K.D.J. acknowledges funding from the MARIE SKŁODOWSKA-CURIE Individual Fellowship under REA grant agreement No. 661416 (SiPhoN). K.D.J. and R.T. acknowledge the COST Action MP1403, supported by COST (European Cooperation in Science and Technology). A.R. acknowledges funding from the Austrian Science Fund (FWF): P 29603. We acknowledge Florian Sipek and Matthias Gartner for help with the experimental setup and data evaluation. We further thank Javier Martin-Sanchez and Johannes Wildmann for fruitful discussions as well as Oliver G. Schmidt for providing access to the MBE facility.

\bibliography{references}
\newpage
\section*{TOC Figure}
\begin{figure}[ht]
\centering
\includegraphics{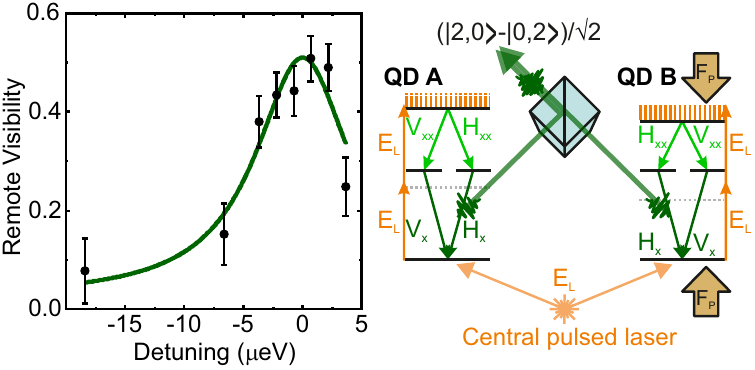}
\label{fig:TOC}
\end{figure}

\end{document}